\documentclass{ws-ijmpe}
\begin{document}

\markboth{A. B. Balantekin and  H. Y\"{u}ksel}
{Neutrino Physics and Nuclear Axial Two-Body Interactions}

\catchline{}{}{}{}{}

\title{NEUTRINO PHYSICS\\ AND\\ NUCLEAR AXIAL TWO-BODY INTERACTIONS}

\author{\footnotesize  A. B. BALANTEKIN}
\address{Department of Physics, University of Wisconsin\\
1150 University Avenue\\
Madison, Wisconsin 53706 USA\\
baha@physics.wisc.edu }

\author{\footnotesize  H. Y\"{U}KSEL}
\address{Department of Physics, University of Wisconsin\\
1150 University Avenue\\
Madison, Wisconsin 53706 USA\\
yuksel@physics.wisc.edu}

\maketitle

\begin{history}
\received{\today}
\end{history}

\begin{abstract}
We consider the counter-term describing isoscalar 
axial
two-body currents in the nucleon-nucleon interaction, $L_{1A}$, in the
effective field theory approach. We determine this quantity using the
solar neutrino data. We investigate the variation of $L_{1A}$ when different
sets of data are used. 
\end{abstract}


\section{Introduction}

Few body reactions play an important role in astrophysics and cosmology. 
In recent years nuclear effective field theories were developed for few 
nucleon systems.\cite{Kaplan:1998sz} The question we wish to address
here is if astrophysical data (in particular solar neutrino data) can
be used to constrain effective field theory description of nuclear
reactions. 

The goal of effective field theories is to find an appropriate way to 
integrate over the undesired degrees of freedom. For example one may wish 
to write an effective theory of photon interactions by integrating over 
the charged elementary particles in quantum electrodynamics. To represent 
the photon-photon interaction one may introduce a point interaction of the 
photons instead of the square box diagram with four external photon lines 
and charged-particles circulating in the loop integral. However, such a 
recipe produces a divergent diagram when we go to the next order  
with one photon loop in the effective theory (the equivalent diagram 
with two charged particle loops in the original theory, quantum 
electrodynamics, is normalizable). To circumvent this problem one
introduces counter terms in the effective theory to cancel the
infinities. Such counter terms should of course be consistent with the
symmetries of the original theory.  

The effective field theories can be applied to the neutrino-deuteron 
reactions measured at the Sudbury Neutrino Observatory (SNO) 
\begin{equation}
\label{1}
\nu_e + d \rightarrow e^- + p + p ,
\end{equation}
\begin{equation}
\label{2}
\nu_x + d \rightarrow \nu_x + p + n .
\end{equation}
At low energies below the pion threshold, the $^3S_1 \rightarrow ^3S_0$ 
transition dominates these reactions and one only needs the
coefficient of the two-body counter term, so called
$L_{1A}$.\cite{Butler:1999sv,Butler:2000zp,Ando:2002pv}
This term can be obtained by 
comparing the cross section $ \sigma (E) = \sigma_0 (E) + L_{1A}
\sigma_1 (E)$ with either cross sections calculated using other, more
traditional, nuclear physics approaches
\cite{Ying:1991tf,Doi:tm,Nakamura:2000vp} or with direct 
measurements. Note that once the quantity $L_{1A}$ is determined one can 
easily calculate related two-body reactions such as
\begin{equation}
\label{3}
p + p \rightarrow d + \nu_e + e^+.
\end{equation}
The reaction in Eq. (\ref{3}) is impossible to measure for the very-low 
energies in the solar core. For that reason the process of determining 
$L_{1A}$ was dubbed ``Calibrating the Sun'' by
Holstein.\cite{Holstein:2000pb}  
Note that naive dimensional analysis predicts a value of 
$|L_{1A}| \sim 6$ fm$^3$ when the renormalization scale is set to the
muon mass. Since the value of $L_{1A}$ depends on the renormalization
scale, this dimensional estimate cannot be used at lower energies.

\section{Extraction of $L_{1A}$}

In our calculations we used the neutrino cross sections given in
Refs. 2 and 3. The details of
the analysis method and procedures used to calculate the measured 
neutrino rates
and spectra are described in Refs. 9 and 
10. To calculate the MSW survival probabilities
we used the neutrino spectra and solar electron density profile
given by the Standard Solar Model of Bahcall and
collaborators.\cite{Bahcall:2000nu} In the calculations presented in
this section we 
took the mixing angle $\theta_{13}$ to be zero. What we present below
is an improved version of our analysis in Ref. 12. 

\begin{figure}[ht]
\vspace*{8pt}
\centerline{\psfig{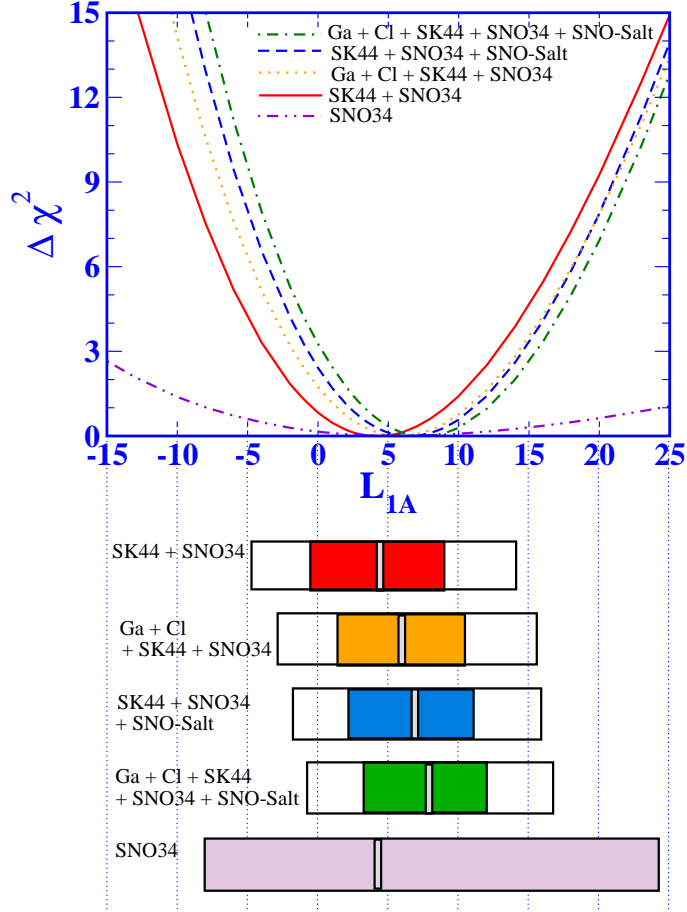}}
\vspace*{8pt}
\caption{
\label{fig:l1afin1}
Best fit values and $1\sigma$, $2\sigma$ bounds of  $ L_{1A} $ obtained 
from various data sets. The upper panel shows the variation of
$\Delta \chi^2$ with $L_{1A}$. The lower panel illustrates $1 \sigma$
(dark areas) and $2 \sigma$ (white areas in the boxes) bounds on
$L_{1A}$.  
For the bound obtained using only the 34 bin
charged-current data from the SNO experiment we show just the
$1\sigma$ limit.
}
\end{figure}

We calculate $\chi^2$ marginalized over the neutrino parameters
$\delta m_{12}^2$ and $\theta_{12}$. 
In Figure \ref{fig:l1afin1} we present the quantity
$\Delta \chi^2 = \chi^2 - \chi^2_{\rm
min}$  calculated as a function of $L_{1A} $. In this figure
$\Delta \chi^2$ is projected only on one parameter ($ L_{1A}$)
so that $n-\sigma$ bounds on it are given by  $\Delta \chi^2 =
n^2$. In our global fit we used 
93 data points from solar and reactor neutrino experiments; namely
the  total rate of the chlorine
experiment (Homestake\cite{Cleveland:nv}), the average rate of
the gallium experiments (SAGE\cite{Abdurashitov:2002nt}, 
GALLEX\cite{Hampel:1998xg}, GNO\cite{Altmann:2000ft}), 44 data
points from the SuperKamiokande (SK) 
zenith-angle-spectrum,\cite{Fukuda:2001nj} 
34 data points from the SNO 
day-night-spectrum,\cite{Ahmad:2001an} the neutral current flux 
measurement at SNO using
dissolved salt\cite{Ahmed:2003kj} and 13 data points from the KamLAND
spectrum\cite{Eguchi:2002dm}. In Figure \ref{fig:l1afin1} we show the
$L_{1A}$ values obtained using only the solar neutrino data from SNO
as well as various combinations of other experiments. Clearly the
$\chi^2$ 
minimum is almost the same in all these cases. In all these cases we
obtain a best fit value of $L_{1A}$ around 5 fm$^3$. It should be
noted that one and two sigma bounds on $L_{1A}$ get significantly
reduced when, in addition to SNO, one includes other solar neutrino
data as well. The $L_{1A}$ values change between 4 and 8 fm$^3$ with a
one-sigma error of 5 fm$^3$. 

\begin{figure}[ht]
\vspace*{8pt}
\centerline{\psfig{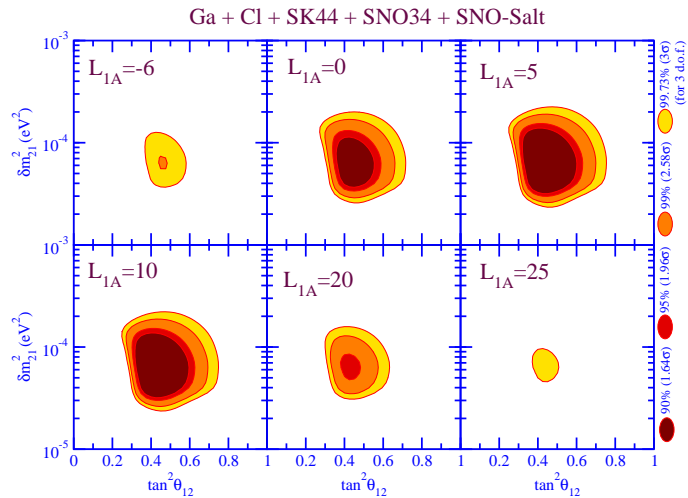}}
\vspace*{8pt}
\caption{\label{fig:l1afin2}
The change in the allowed region of the mixing parameter space
using only the solar neutrino data as a function of
$ L_{1A} $. In the calculations leading
to this figure the neutrino mixing angle $\theta_{13}$ is taken
to be zero. The shaded
areas corresponds to 90 \% , 95 \% ,
99 \% , and 99.73 \% confidence
levels.
}
\end{figure}

\begin{figure}[ht]
\vspace*{8pt}
\centerline{\psfig{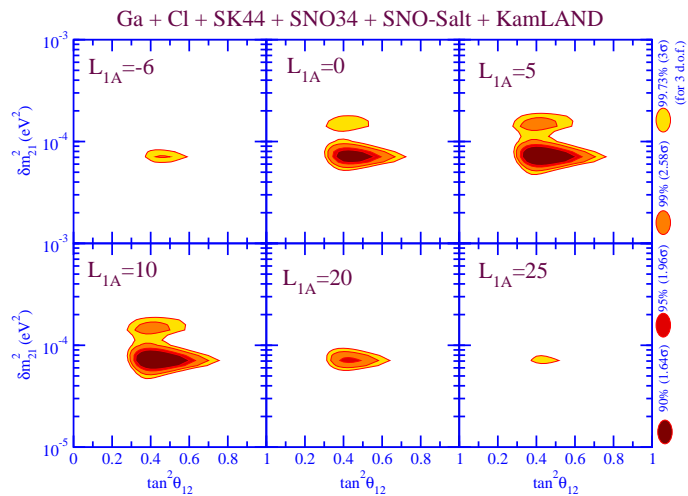}}
\vspace*{8pt}
\caption{\label{fig:l1afin2-kl}
\label{fig:l1afin2b}
The change in the allowed region of the mixing parameter space
using combined solar neutrino data and KamLAND as a function of
$ L_{1A} $. In the calculations leading
to this figure the neutrino mixing angle $\theta_{13}$ is taken
to be zero. The shaded
areas corresponds to 90 \% , 95 \% ,
99 \% , and 99.73 \% confidence
levels.
}
\end{figure}

The dependence of the extracted neutrino parameters on the value of
$L_{1A} $ is not very strong. We show how the neutrino parameter space
changes with $L_{1A} $ in Figures \ref{fig:l1afin2}  and
\ref{fig:l1afin2b}. The analysis presented in Figure \ref{fig:l1afin2}
uses only the solar neutrino data as input whereas that presented in 
Figure \ref{fig:l1afin2b} uses both the solar neutrino data and
results from the KamLAND reactor neutrino measurements. 

\section{Exploring the effects of $\theta_{13}$}

\begin{figure}[ht]
\vspace*{8pt}
\centerline{\psfig{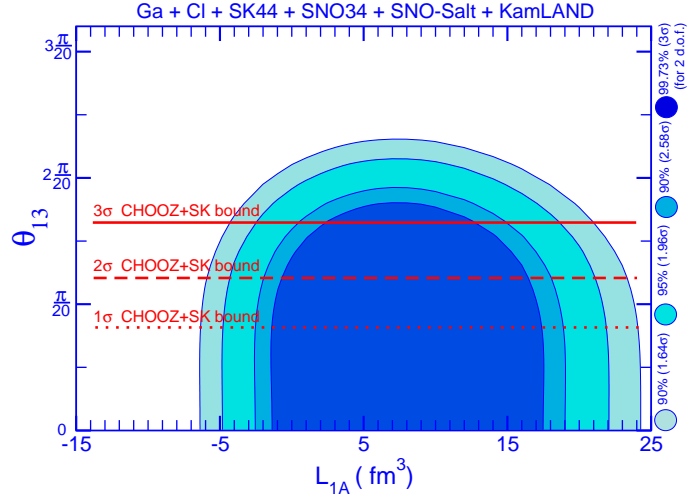}} 
\vspace*{8pt}
\caption{\label{fig:l1afin3}
Allowed parameter space for $L_{1A}$ vs $\theta_{13}$ when 
$\chi^2$ is marginalized over
$\theta_{12}$ and $ \delta m_{12}^2 $. All
solar neutrino experiments along with the KamLAND experiment considered.
The shaded areas are the 90 \%  95 \% ,
99 \% , and 99.73 \%  confidence
levels. Horizontal lines show $\theta_{13}$ bounds from CHOOZ + SK.
}
\end{figure}

One of the open questions in neutrino physics is understanding the
role of mixing between the first and third flavor generations,
$\theta_{13}$. Since both of the quantities $\theta_{13}$ and $L_{1A}$
are rather small one can investigate if the
uncertainties coming from the lack of knowledge of $\theta_{13}$
and the counter-term $L_{1A} $ are comparable. In the limiting case
of small $ \cos {\theta_{13}}  $ and $ \delta m_{31}^2
\gg \delta m_{21}^2 $, which seems to be satisfied by the
measured neutrino properties, it is possible to incorporate the
effects of $\theta_{13}$ rather easily. In this limit the 
charged-current counting
rate at SNO can be linearized in
$\cos^4{\theta_{13}}$:\cite{Balantekin:2003ep} 
\begin{equation}
\mathrm{Count} \, \mathrm{Rate} \sim
A + B \, (1 - \cos^4{\theta_{13}} ), 
\end{equation}
where the parameters $A$ and $B$ are independent of $\theta_{13}$. 
The neutral- and charged-current counting rates linearly depend on
$L_{1A} $ while elastic scattering rate does not.
Conversely the charged-current and elastic scattering rates
linearly depend on $\cos^4{\theta_{13}}$ while the neutral-current
rate does not. We show the allowed $\theta_{13}$
and $L_{1A} $ parameter space in Figure \ref{fig:l1afin3} 
when $\theta_{12}$ and $ \delta
m_{12}^2 $ are taken to give the minimum $\chi^2$ values to 
fit the data.

Of course there are other experiments that limit the value of 
$\theta_{13}$. The CHOOZ\cite{Apollonio:2002gd} and Palo
Verde\cite{Boehm:2001ik} experiments limit $\sin^2 2\theta_{13}$ to be
less than 0.19 at 90\% C.L. for $\delta m^2_{\rm atmos} = 0.002$
eV$^2$. Data from the K2K experiment\cite{Ahn:2004te} 
provides a limit of $\sin^2 2\theta_{13} < 0.3$. These limits are
consistent with SK atmospheric neutrino data.\cite{Nakaya:2002ki} We
also show these bounds in Figure \ref{fig:l1afin3}. Note that future
data from SNO and KamLAND may further limit the value of  
$\theta_{13}$.\cite{Balantekin:2004hi}

\section{Conclusions}

Several other authors tried to estimate the value of $L_{1A}$. Using
SNO and SK data Chen, Heeger, and Robertson found\cite{Chen:2002pv} 
\begin{equation}
L_{1A}  = 4.0 \pm 6.3 \: \mathrm{fm}^3 .
\end{equation}
As we mentioned in the Introduction, the $pp$ fusion cross section
also depends on $L_{1A}$. State of the art calculations of this cross
section implies a value of\cite{Schiavilla:1998je}   
\begin{equation}
L_{1A}  = 4.2 \pm 2.4   \: \mathrm{fm}^3 
\end{equation}
in third order of power counting. (This is the same order in which the
neutrino-deuteron cross-section of Refs. 2 and 3 are calculated).
              
Helioseismic observation of the
pressure-mode oscillations of the 
Sun can be used to put constraints on various inputs into the
Standard Solar Model, in particular the $pp$ fusion cross section.
Helioseismology gives a limit of\cite{Brown:2002ih} 
\begin{equation}
L_{1A}  = 4.8 \pm 6.7  \: \mathrm{fm}^3
\end{equation}
in the third order. Finally one can present a constraint on $L_{1,A}$
imposed by existing reactor antineutrino-deuteron breakup
data,\cite{Butler:2002cw} which yields 
\begin{equation}
L_{1A}  = 3.6 \pm 5.5  \: \mathrm{fm}^3 .
\end{equation}
Various values of $L_{1,A}$ we obtained using different data sets are
rather comparable to the values listed above. 

The uncertainties in the neutrino-deuteron breakup
cross-sections at low energies are dominated by the isovector axial
two-body current parametrized by $L_{1,A}$. However  the contribution
of the uncertainty in 
$L_{1A} $ to the analysis and interpretation of the SNO data is
rather small.\cite{Butler:2002cw,Balantekin:2003ep} The effect 
of this uncertainty is even smaller than
effects of a value of $\theta_{13}$ near its currently allowed maximum
or than effects
of possible solar density fluctuations \cite{Balantekin:2003qm}.

\section*{Acknowledgments}

We would like to express our gratitude to I.H. Duru and other 
personnel of the Feza
G\"ursey Institute for helping us to put together a very successful
conference. 
We also thank the members of the International Advisory Committee,
in particular C. Johnson, for their efforts. 
This  work   was supported in  part  by   the  U.S.  National  Science
Foundation Grant No. PHY-0244384 at the  University
of  Wisconsin, and  in  part by  the  University of Wisconsin Research
Committee   with  funds  granted by    the  Wisconsin Alumni  Research
Foundation.

\end{document}